\RequirePackage{fix-cm}
\documentclass[twocolumn]{svjour3}          
\smartqed  
\usepackage{graphicx}
\usepackage{amsmath,amssymb}
\begin{document}

\title{Resonant-state expansion of the Green's function of open quantum systems
}


\author{Naomichi Hatano         \and
        Gonzalo Ordonez 
}


\institute{N. Hatano \at
              Institute of Industrial Science, University of Tokyo, 4-6-1 Komaba, Meguro, Tokyo 153-8505, Japan \\
              Tel.: +81-3-5452-6154\\
              Fax: +81-3-5452-6155\\
              \email{hatano@iis.u-tokyo.ac.jp}           
           \and
           G. Ordonez \at
              Department of Physics, Butler University, 4600 Sunset Ave., 
Indianapolis, IN 46208, USA
}

\date{Received: date / Accepted: date}

\maketitle

\begin{abstract}
Our series of recent work on the transmission coefficient of open quantum systems in one dimension will be reviewed.
The transmission coefficient is equivalent to the conductance of a quantum dot connected to leads of quantum wires.
We will show that the transmission coefficient is given by a sum over all discrete eigenstates without a background integral.
An apparent ``background" is in fact not a background but generated by tails of various resonance peaks.
By using the expression, we will show that the Fano asymmetry of a resonance peak is caused by the interference between various discrete eigenstates.
In particular, an unstable resonance can strongly skew the peak of a nearby resonance.
\end{abstract}

\section{Introduction}

Open quantum systems have renewed researchers' interest repeatedly.
The first rise of the interest was obviously initiated by quantum scattering theory of atoms and nuclei.
Indeed, a quantum scatterer embedded in an infinitely wide space is an open quantum system, although it might have not been termed so.
The existence of resonant states with complex eigenvalues suggested the non-Hermiticity of open quantum systems.

One of the recent rises of the interest in open quantum systems was perhaps triggered by the development of nanotechnology.
Mesoscopic objects such as quantum dots fabricated in semiconductor heterostructures are, at low temperatures, an ideal playground of quantum mechanics.
Quantum dots, when connected to electron reservoirs through leads, are indeed open quantum systems.
As we will review in the next section, the conductance between electron reservoirs is essentially the transmission coefficient of the quantum dot and exhibits various forms of resonance peaks that are common to other open quantum systems.

Many measurements of the conductance of quantum dots connected to quantum wires have motivated us to carry out a series of recent work~\cite{Sasada05,Hatano07,Hatano09,Sasada10} on the transmission coefficient of open quantum systems in one dimension.
The main purpose of the present article is to review the work.
We will emphasize the following two points:
\begin{enumerate}
\renewcommand{\labelenumi}{(\roman{enumi})}
\setlength{\labelsep}{1em}
\settowidth{\labelwidth}{(ii)}
\setlength{\itemindent}{1em}
\item The transmission coefficient is given by a sum over all discrete eigenstates without a background integral.
An apparent ``background" is in fact not a background but generated by tails of various resonance peaks.
\item The Fano asymmetry of a resonance peak is caused by the interference between various discrete eigenstates.
In particular, an unstable resonance can strongly skew the peak of a nearby resonance.
\end{enumerate}

The paper is organized as follows.
We will first argue in Sec.~\ref{sec2} the physical significance of the resonance, particularly its dissipative feature from the viewpoint of the conductance of a quantum dot.
We then will review in Sec.~\ref{sec3} some known facts on resonant states and other discrete eigenstates.
Section~\ref{sec4} features the first point (i) of our work, whereas Sec.~\ref{sec5} features the second point (ii).
The final section will be devoted to a summary.

\section{Landauer formula}
\label{sec2}

\begin{figure}
\centering
\includegraphics[width=0.8\columnwidth]{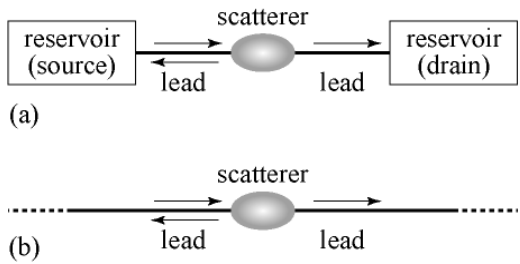}
\caption{(a) A quantum scatterer is connected to two electron reservoirs through finite leads. The conductance $g$ between the source and the drain is a physical observable.
(b) A quantum scatterer sits on the infinitely long leads, which constitutes an open quantum system.
The transmitted and reflected waves dissipate into the right and left infinities, respectively, and never come back into the scatterer.}
\label{fig1}
\end{figure}

The starting point of our work is the Landauer formula of the conductance of a mesoscopic quantum system.
The conductance $g$ between the two electron reservoirs, namely the source and the drain, shown in Fig.~\ref{fig1}(a) is given by~\cite{Datta95}
\begin{align}\label{eq10}
g(E)=\frac{2e^2}{h}T(E),
\end{align}
where $T$ is the transmission coefficient of the quantum scatterer shown in Fig.~\ref{fig1}(b).
We argue here why a physical quantity in the situation of the finite system in Fig.~\ref{fig1}(a) is related to one in the situation of the infinite system in Fig.~\ref{fig1}(b).

First of all, the finite conductance~\eqref{eq10} means a finite resistance $1/g$, which in turn means a dissipation.
Where does this dissipation take place?
It cannot occur around the quantum scatterer nor in the leads, because we assume purely quantum-mechanical propagation of electrons there.
In fact, the dissipation takes place in the electron reservoirs, or more precisely, their contact with the leads;
the quantum coherence of the electrons that come from the leads into the reservoirs is completely lost before the electrons return to the leads onto the quantum scatterer again.
The process of the loss of the coherence yields the dissipation in the situation in Fig.~\ref{fig1}(a).

This loss of the coherence is mimicked in the situation in Fig.~\ref{fig1}(b).
Electrons that are scattered by the quantum scatterer go away into the right and left infinities and never come back into the system.
Therefore, electrons after the scattering never correlate with other electrons before the scattering.
This non-correlation is equivalent to the loss of the coherence in the electron reservoirs.

The above argument makes us notice that the infinite system in Fig.~\ref{fig1}(b), namely the open quantum system, does have a dissipation, the dissipation of particles into the infinite leads.
In fact, this dissipation of the open quantum system makes the system Hamiltonian non-Hermitian and is precisely described by its resonant states.
This was elaborated in Ref.~\cite{Hatano07}, which showed that:
\begin{enumerate}
\renewcommand{\labelenumi}{(\roman{enumi})}
\setlength{\labelsep}{1em}
\settowidth{\labelwidth}{(ii)}
\setlength{\itemindent}{1em}
\item the non-Hermiticity of the open quantum system is caused by particle dissipation out into the infinite leads;
\item the resonant states with complex eigenvalues are eigenstates of the system with the boundary conditions of outgoing waves only.
\end{enumerate}

Hereafter, we will set aside the situation in Fig.~\ref{fig1}(a) and focus on the resonant states in the situation in Fig.~\ref{fig1}(b).
However, the readers should always remember that the dissipation caused by the resonant states is a physical observable in the form of the conductance.

\section{Resonant and other discrete eigenstates}
\label{sec3}

\begin{figure}
\centering
\includegraphics[width=0.8\columnwidth]{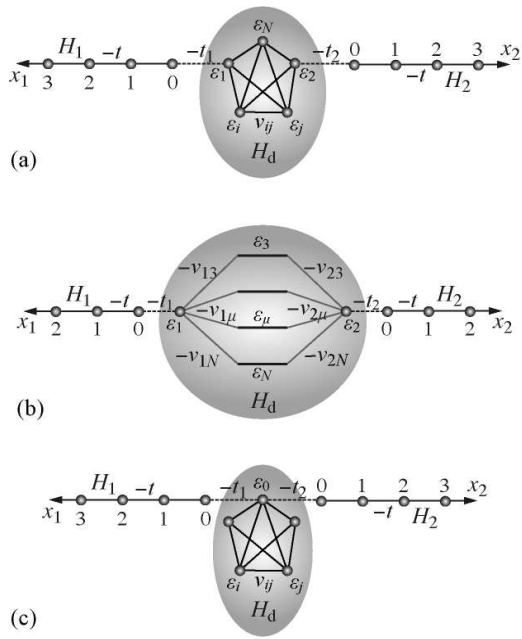}
\caption{(a) A schematic view of the system that we consider in the present work.
(b) The dot Hamiltonian is partially diagonalized. This system is included in the system in (a).
(c) The system with two leads attached to the same site $0$.}
\label{fig2}
\end{figure}

The system that we consider hereafter is schematically shown in Fig.~\ref{fig2}(a).
The system consists of the dot Hamiltonian $H_\textrm{d}$, the lead Hamiltonians $H_1$ and $H_2$ and the hopping between the dot and a lead.
The dot Hamiltonian is a tight-binding system of $N$ sites with arbitrary hopping amplitudes and arbitrary on-site potentials.
Each lead Hamiltonian is a semi-infinite tight-binding system with a uniform hopping amplitude $t$ and has the dispersion relation $E(k)=-2t\cos k$.
The contact sites, to which the leads are connected, are designated as the sites $1$ and $2$.
The respective coupling amplitude between the dot and a lead, $t_1$ and $t_2$, can be arbitrary.
The system is general enough to include the system in Fig.~\ref{fig2}(b), where the dot Hamiltonian is partially diagonalized to a number of energy levels.

Before going into the main part of our work, let us briefly review known facts on resonant and other discrete eigenstates;
see Ref.~\cite{Hatano07} for details.
The resonant state can be defined as an eigenstate of the stationary Schr\"{o}dinger equation with boundary conditions of out-going waves only:
\begin{align}\label{eq15}
\lim_{x\to\pm\infty}\psi(x)=e^{ik|x|},
\end{align}
which is called the Siegert condition~\cite{Siegert1939}.
In the case $\mathop{\mathrm{Re}}k\neq 0$, the state does not conserve the particle number in the naive sense.
This leads to the non-Hermiticity of the Hamiltonian operator~\cite{Hatano07}.
The Hamiltonian then can produce a complex eigenvalue.

Such an eigenstate with a complex eigenvalue does not belong to the Hilbert space.
The seemingly Hermitian Hamiltonian can be non-Hermitian outside the Hilbert space.
Indeed, the corresponding eigenfunction diverges in the limit $|x|\to\infty$.
The complex eigen-wave-number $k_n$, which is related to the complex energy eigenvalue $E_n$ through the dispersion relation $E_n=-2t\cos k_n$ has a negative imaginary part and causes the divergence in Eq.~\eqref{eq15}.
The spatially diverging wave function is obviously outside the Hilbert space and hence can accommodate a complex eigenvalue.
We can also show that the spatial divergence is physically necessary for particle-number conservation in an extended sense~\cite{Hatano07,Hatano09}.
When we count the number of particles appropriately, the spatial divergence is cancelled by the temporal decay and thereby the number of particles is conserved.

For tight-binding systems such as the present one, there is an efficient method of finding the eigenstates that satisfy the Siegert condition~\eqref{eq15}.
The method is sometimes referred to as the method of the effective Hamiltonian.
See Ref.~\cite{Sasada10} for details.
We can also show that there are generally $2N$ eigenstates with discrete eigenvalues for the dot with $N$ sites.

\begin{figure}
\centering
\includegraphics[width=0.7\columnwidth]{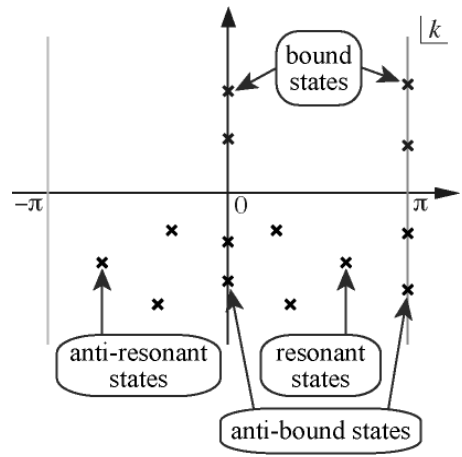}
\caption{The crosses on the positive imaginary axis as well as on the positive part of the $k=\pi$ line designate the bound states.
The crosses in the fourth quadrant designate the resonant states, while the crosses in the third quadrant designate the anti-resonant states.
Each resonant state has an anti-resonant state as a partner.
Their locations are mirror images with respect to the imaginary axis.
The crosses on the negative real axis as well as on the negative part of the $k=\pi$ line designate the anti-bound states.
Note that the $k=-\pi$ line is identified with the $k=\pi$ line because of the lattice periodicity.}
\label{fig3}
\end{figure}

Figure~\ref{fig3} shows the classification of the discrete eigenstates in terms of their locations in the complex wave-number plane.
Because of the lattice periodicity of the tight-binding leads, the wave-number plane is restricted to the Brillouin zone $-\pi\leq\mathop{\mathrm{Re}}k\leq\pi$ and the line $k=-\pi$ is identified with the line $k=\pi$.
The positive parts of the imaginary axis and the $k=\pi$ line have bound states.
A positive imaginary part of the eigen-wave-number indeed makes the wave function~\eqref{eq15} decay exponentially in space.
The bound states on the $k=\pi$ line do not exist for problems in the continuum space;
they are characteristic to lattice problems.

The resonant states are in the fourth quadrant of the complex wave-number plane.
The negative imaginary part of the eigen-wave-number makes the wave function spatially divergent and pushes it out of the Hilbert space.
The positive real part of the eigen-wave-number indicates a particle flow away from the scatterer into the infinite leads.

Each resonant state has a partner in the third quadrant, which is referred to as an anti-resonant state.
(Note, however, that other authors sometimes use the term anti-resonance to refer to a resonance dip, not a peak.)
The positions of a resonant state and the corresponding anti-resonant state are symmetric with respect to the imaginary axis.
The anti-resonant state is the time-reversal of the resonant state.
The negative real part of the eigen-wave-number indicates a particle flow into the scatterer.

Depending on the system parameters, there sometimes exist discrete states on the negative parts of the imaginary axis and the $k=\pi$ line.
These states are called anti-bound states.
An anti-bound state can arise when a bound state moves from the upper half plane to the lower half plane.
Two anti-bound states can be also born when a resonant state and the corresponding state collide on the imaginary axis.

Other than the above discrete eigenstates, there are scattering states $\psi_k$ that form a continuum on the real $k$ axis.
It has been proved that the bound states and the continuum of the scattering states constitute a resolution of unity~\cite{Newton1961},
\begin{align}\label{eq18}
1=\sum_{n\in \textrm{bound}}|\psi_n\rangle\langle\psi_n|+\int_{-\pi}^\pi|\psi_k\rangle\langle\psi_k|dk.
\end{align}

\section{Resonant-state expansion of the Green's function}
\label{sec4}

Let us come back to the transmission coefficient of the quantum scatterer in Fig.~\ref{fig2}.
The transmission coefficient in Eq.~\eqref{eq10} is known to be written in the form~\cite{Datta95}
\begin{align}\label{eq20}
T=\mathop{\mathrm{Tr}}G^\textrm{R}\Gamma_{11}G^\textrm{A}\Gamma_{22},
\end{align}
where $G^\textrm{R}$ and $G^\textrm{A}$ are $N$-by-$N$ matrices whose elements are the retarded and advanced Green's functions between $i$th and $j$th sites of the dot.
The matrix $\Gamma$ is also an $N$-by-$N$ matrix of the form
\begin{align}\label{eq30}
\Gamma=\frac{\sqrt{4t^2-E^2}}{t^2}
\begin{pmatrix}
{t_1}^2 & 0 & 0 & \cdots \\
0 & {t_2}^2 & 0 & \cdots \\
0 & 0 & 0 & \cdots \\
\vdots & \vdots & \vdots & \ddots
\end{pmatrix},
\end{align}
where the first column and row correspond to the contact site $1$ while the second column and row correspond to the contact site $2$.

We have rewritten Eq.~\eqref{eq20} in the form~\cite{Sasada10}
\begin{align}\label{eq40}
T=\Gamma_{11}\Lambda_{12}\Gamma_{22}\Lambda_{21}
\frac{-(D-4)\pm\sqrt{(D+4)^2-4T^2}}{2(T^2-4D)},
\end{align}
where
\begin{align}
\Lambda=G^\textrm{R}+G^\textrm{A}
\end{align}
and
\begin{align}
T=\mathop{\textrm{Tr}}\check{\Gamma}\check{\Lambda},\quad
D=\det\check{\Gamma}\check{\Lambda}
\end{align}
with $\check{\Gamma}$ and $\check{\Lambda}$ being top-left two-by-two matrices cut out of the $N$-by-$N$ matrices $\Gamma$ and $\Lambda$, respectively.

Incidentally, the form~\eqref{eq40} reduces to a much simpler form when the two leads are attached to the same site of the dot as shown in Fig.~\ref{fig2}(c)~\cite{Sasada10}:
\begin{align}\label{eq65}
T=\left(\frac{t_1t_2}{{t_1}^2+{t_2}^2}\right)^2\left[2\pm\sqrt{4-\left(\Gamma_{00}\Lambda_{00}\right)^2}\right],
\end{align}
Here we denoted the contact site as the site $0$ and
\begin{align}
\Gamma_{00}=\frac{\sqrt{4t^2-E^2}}{t^2}\left({t_1}^2+{t_2}^2\right).
\end{align}
We will use this form in the next section for simplicity when we consider interferences that cause the Fano asymmetry.

The rewriting in the form~\eqref{eq40} is seemingly a complication of Eq.~\eqref{eq20}, but the purpose is to use the matrix $\Lambda=G^\textrm{R}+G^\textrm{A}$ instead of using $G^\textrm{R}$ and $G^\textrm{A}$ separately.
This is because we have found the resonant-state expansion of the matrix $\Lambda$ in the following form~\cite{Sasada10}
\begin{align}\label{eq70}
\Lambda=\sum_{n}\frac{|\psi_n\rangle\langle\tilde{\psi}_n|}{E-E_n},
\end{align}
where $|\psi_n\rangle$ and $\langle\tilde{\psi}_n|$ are the right- and left-eigenvectors with the eigenvalue $E_n$ of each discrete eigenstate of the present open quantum system.
Note that $E_n$ is generally complex for resonant and anti-resonant states.

\begin{figure}
\centering
\includegraphics[width=0.7\columnwidth]{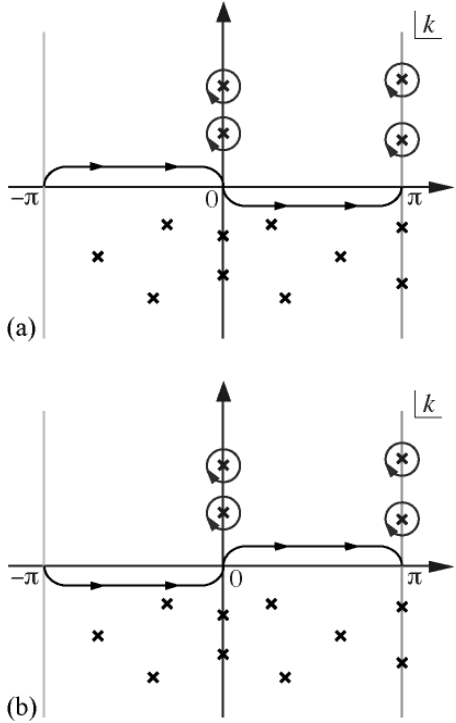}
\caption{The integration contours for (a) $G^\textrm{R}$ and (b) $G^\textrm{A}$.}
\label{fig4}
\end{figure}

The important feature of the resonant-state expansion~\eqref{eq70} is the fact that it has no background integral.
Such an expansion is indeed quite rare.
As far as we know, the only other expansion is the one with respect to the wave number~\cite{Tolstikhin01,Ostrovsky05,Klaiman10}.
In all other studies, some forms of background integral remain because $G^\textrm{R}$ and $G^\textrm{A}$ are used separately, not in the form of $\Lambda=G^\textrm{R}+G^\textrm{A}$.
Because of the resolution of unity~\eqref{eq18}, the Green's function is given by
\begin{align}
G^\textrm{R/A}=\sum_{\underset{\textrm{\scriptsize states}}{n:\textrm{ bound}}}\frac{|\psi_n\rangle\langle\psi_n|}{E-E_n}+\int_{-\pi}^\pi\frac{|\psi_k\rangle\langle\psi_k|}{E-E_k\mp i\eta\mathop{\mathrm{sgn}}k}dk,
\end{align}
where $E_n$, the bound-state energies, and $E_k$, the scattering-state energies, are both real; $\eta$ is infinitesimal; and $\mathop{\mathrm{sgn}}k$ is the sign of $k$.
The contours of these integrals for $G^\textrm{R}$ and $G^\textrm{A}$ are schematically shown in Fig.~\ref{fig4}.
Some of the resonant states in the fourth quadrant and some of the anti-resonant states in the third quadrant can be taken into account by modifying the integration contours.
No matter how modified, however, the integral remains, which constitutes the background integral.

\begin{figure}
\centering
\includegraphics[width=0.7\columnwidth]{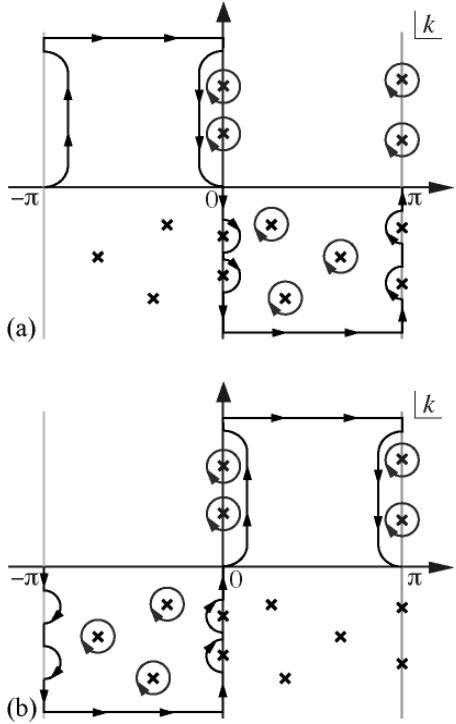}
\caption{The modified integration contours for (a) $G^\textrm{R}$ and (b) $G^\textrm{A}$.
The former contains the resonant states as well as the half contribution of the anti-bound states.
The latter contains the anti-resonant states as well as the half contribution of the anti-bound states.}
\label{fig5}
\end{figure}

The essential point of our expansion~\eqref{eq70} is first to modify the contours as shown in Fig.~\ref{fig5} and then to sum up the two.
Then the contours on the real axis as well as on the $k=\pi$ line are cancelled out.
(Note here that the $k=-\pi$ line is identified with the $k=\pi$ line because of the lattice periodicity.)
We also proved that the upper and lower horizontal paths give zero contributions in the limit $|\mathop{\mathrm{Im}}k|\to\infty$.
We thereby end up with the contributions of all of the discrete eigenstates only, no more integrals, as in Eq.~\eqref{eq70}.

The expansion~\eqref{eq70} without the background integral shows that there is in fact no background integral in the conductance profile~\eqref{eq10}.
We often see explanations of the conductance profile (the energy dependence of the conductance) as resonance peaks with a background.
Our expansion clearly claims that the ``background" is in fact not a background, but is formed by tails of all other peaks.

\section{Interference of resonant states and the Fano asymmetric peaks}
\label{sec5}

We now discuss the origin of the Fano asymmetric peaks of the conductance profile in terms of the interference between discrete eigenstates.
We can show in Eq.~\eqref{eq40} and more clearly in the simpler form~\eqref{eq65} that the conductance profile contains
\begin{align}\label{eq120}
\left(\Lambda_{ij}\right)^2
&=\sum_n\left(\frac{\langle i|\psi_n\rangle\langle\tilde{\psi}_n|j\rangle}{E-E_n}\right)^2
\nonumber\\
&+2\sum_{m<n}\frac{\langle i|\psi_m\rangle\langle\tilde{\psi}_m|j\rangle}{E-E_m}
\frac{\langle i|\psi_n\rangle\langle\tilde{\psi}_n|j\rangle}{E-E_n}.
\end{align}
We showed in Refs.~\cite{Sasada05,Sasada10} that the Fano asymmetry comes from the second line of Eq.~\eqref{eq120}, namely the interference between two discrete states.
We stress here again that the argument does not omit any terms thanks to the fact that the expansion does not contain any background integrals.

\begin{figure}
\centering
\includegraphics[width=0.7\columnwidth]{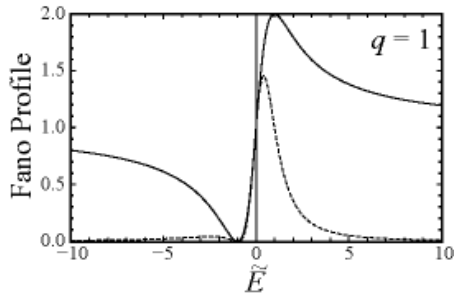}
\caption{The broken curve indicates the profile~\eqref{eq130} of the type~(i).
The solid curve indicates the standard Fano profile~\eqref{eq150} of the types~(ii) and~(iii).
We used the values $q=q'=1$ in plotting these curves.}
\label{fig6}
\end{figure}

The interferences exist between various discrete states as follows:
\begin{enumerate}
\renewcommand{\labelenumi}{(\roman{enumi})}
\setlength{\labelsep}{1em}
\settowidth{\labelwidth}{(iii)}
\setlength{\itemindent}{1em}
\item between a resonant state and the corresponding anti-resonant state;
\item between a resonant-state pair (the pair of a resonant state and the corresponding anti-resonant state) and a bound state or an anti-bound state;
\item between two resonant-state pairs.
\end{enumerate}
We found~\cite{Sasada10} that the first type of the interference, the type (i), produces a form of asymmetry different from Fano's result~\cite{Fano61} (the broken curve in Fig.~\ref{fig6}).
The other two, the types~(ii) and~(iii), follow Fano's line shape (the solid curve in Fig.~\ref{fig6}).

More specifically, the type (i) gives
\begin{align}\label{eq130}
g(E)\simeq\left(\frac{q'+\tilde{E}}{1+\tilde{E}^2}\right)^2,
\end{align}
where $q'$ is the index that specifies the amount of the asymmetry and
\begin{align}
\tilde{E}=\frac{E-\mathop{\textrm{Re}}E_n}{|\mathop{\textrm{Im}}E_n|}
\end{align}
is the energy variable normalized for the resonance $E_n$.
The types~(ii) and~(iii) give a profile that conforms to the original Fano profile
\begin{align}\label{eq150}
g(E)\simeq\frac{\left(q+\tilde{E}\right)^2}{1+\tilde{E}^2},
\end{align}
where $q$ is the original Fano parameter, which specifies the amount of the Fano asymmetry.
We succeeded~\cite{Sasada10} in deriving microscopic expressions of the Fano parameters $q$ and $q'$ from the expansion~\eqref{eq120}.

\begin{figure}
\centering
\includegraphics[width=0.7\columnwidth]{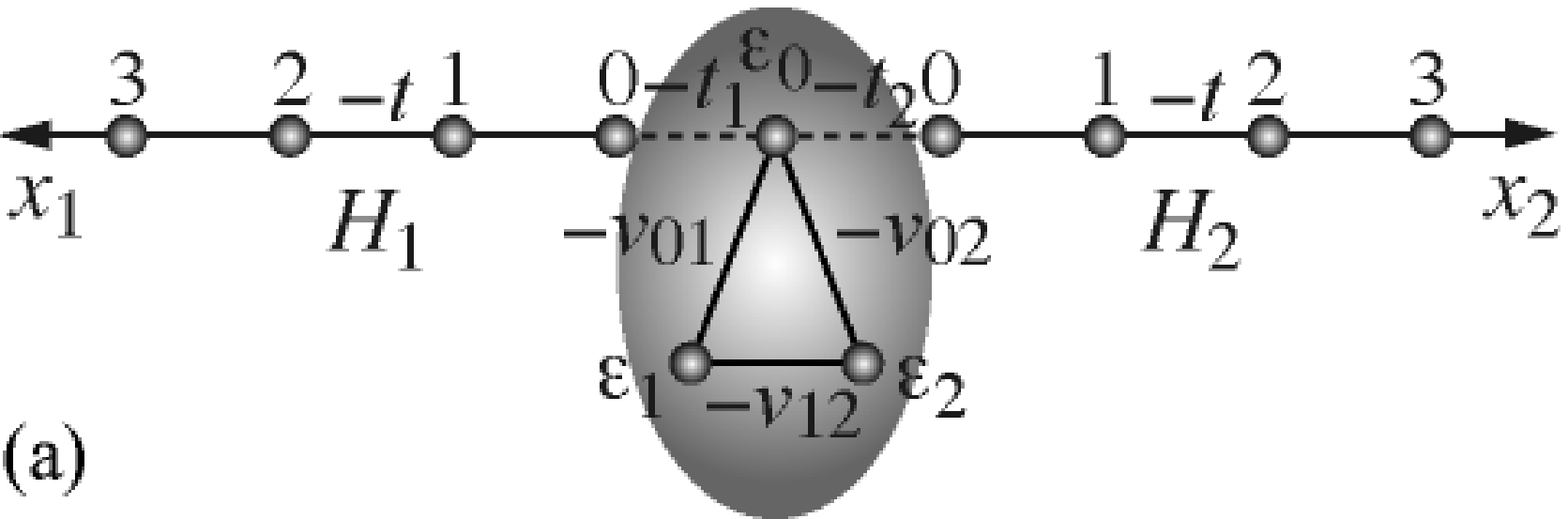}\\
\vspace*{\baselineskip}
\includegraphics[width=0.9\columnwidth]{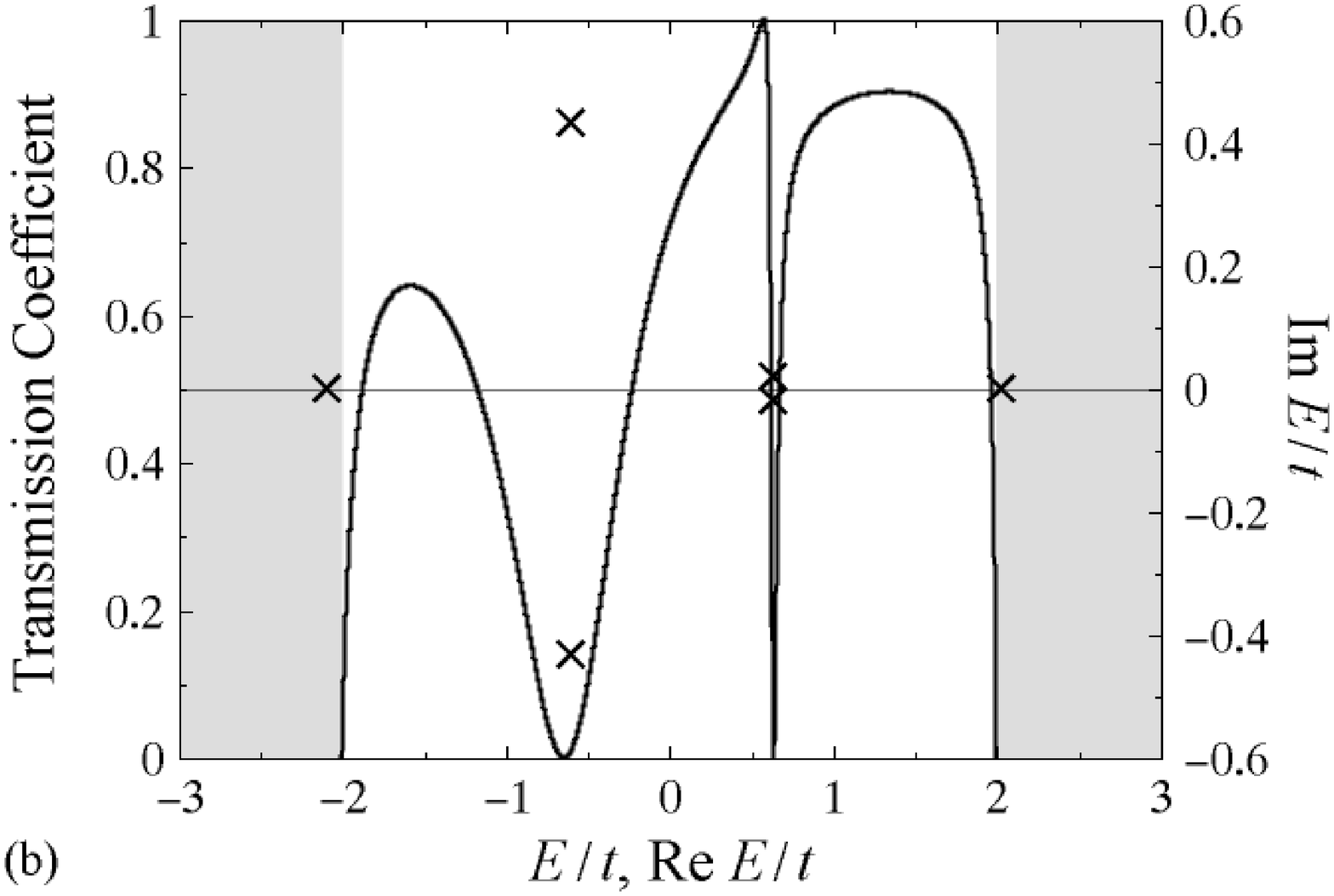}
\caption{(a) A quantum dot with $N=3$.
(b) The left axis indicates the conductance profile, whereas the right axis indicates the locations of the discrete eigenvalues.
The parameter values are as follows: $\varepsilon_0/t=0$; $\varepsilon_1/t=-0.5$; $\varepsilon_2/t=0.5$; $v_{01}/t=0.8$; $v_{02}/t=0.5$; $v_{12}/t=0.4$; $t_1=t_2=t$.}
\label{fig7}
\end{figure}

Let us finally present an interesting example of the conductance profile.
For the system shown in Fig.~\ref{fig7}(a), we obtained the conductance profile in Fig.~\ref{fig7}(b).
This particular system has two bound states (located on the real energy axis on the left and the right of the energy band $-2t\leq E\leq 2t$) and two resonance pairs.
The resonance pair on the left generates a broad, almost symmetric dip in the conductance profile, whereas the resonance pair on the right generates a sharp, very asymmetric Fano peak.
Analysis with the use of the Fano parameter $q$ revealed~\cite{Sasada10} that the Fano asymmetry of the resonance pair on the right is partly caused by the interference between the two resonance pairs.
A more general argument~\cite{Sasada05} indeed showed that, if there are two resonance pair, one of them have a large imaginary part, and the other has a small imaginary part, then the latter resonance pair develops a strong asymmetry.

This example points out the following important fact.
A resonance far from the real axis itself is quite unstable, produces only a broad peak, and hence is generally thought not to contribute to the conductance profile much.
Such a resonance, however, can manifest itself as a strong asymmetry of the resonance peak of a nearby state.
The present quantitative analysis suggests the possibility of detecting a resonance far away from the real axis by means of the Fano asymmetry of a nearby resonance.

\section{Summary}
\label{sec6}

We have reviewed our series of recent work~\cite{Sasada05,Hatano07,Hatano09,Sasada10} on the conductance of a tight-binding quantum dot connected to tight-binding leads.
We have shown for the open quantum system that the conductance profile is given by a sum over all discrete eigenstates without a background integral.
The expression revealed that the Fano asymmetry is caused by interferences between various discrete eigenstates and enabled us to derive microscopic formulas of the Fano parameters.

\section*{Acknowledgments}
The present work is supported by Core Research for Evolutional Science and Technology (CREST) from Japan Science and Technology Agency (JST) as well as by Grant-in-Aid for Scientific Research (B) No.~22340110 from Ministry of Education, Culture, Sports, Science and Technology, Japan.

\bibliographystyle{spphys}       
\bibliography{resonance}   


\end{document}